\documentclass[pre,twocolumn,amsmath,amssymb,showpacs]{revtex4-1}
\usepackage[english]{babel}
\usepackage{graphicx,amstext}
\usepackage{dcolumn}
\usepackage[utf8]{inputenc}
\usepackage[T1]{fontenc}
\usepackage{epstopdf}
\usepackage{bm}

\newcommand{\al}{{\it et al.\@}}
\newcommand{\bq}{\begin{equation}}
\newcommand{\eq}{\end{equation}}
\begin{document}
\title{Effect of magnesium ions on dielectric relaxation in semidilute DNA aqueous solutions}
 
\author{D.\ Grgi\v{c}in}
\homepage{http://real-science.ifs.hr/}
\affiliation{Institut za fiziku, 10000 Zagreb, Croatia}
\author{S.\ Dolanski Babi\'{c}}
\altaffiliation {Permanent address: Department of physics and biophysics, Medical School, University of Zagreb, 10000 Zagreb, Croatia.}
\author{T.\ Ivek }
\affiliation{Institut za fiziku, 10000 Zagreb, Croatia}
\author{S.\ Tomi\'{c}}
\affiliation{Institut za fiziku, 10000 Zagreb, Croatia}
\author{R.\ Podgornik}
\affiliation{Department of Physics, University of Ljubljana and J.\ Stefan Institute, 1000 Ljubljana, Slovenia}
\date{\today}

\pacs{87.15.H-, 82.39.Pj, 77.22.Gm} 
 
\begin{abstract}
The effect of magnesium ion Mg$^{2+}$ on the dielectric relaxation of semidilute DNA aqueous solutions has been studied by means of dielectric spectroscopy in the 100~Hz--100~MHz frequency range. de Gennes--Pfeuty--Dobrynin semidilute solution correlation length is the pertinent fundamental length scale for sufficiently low concentration of added salt, describing the collective properties of Mg-DNA solutions. No relaxation fingerprint of the DNA denaturation bubbles, leading to exposed hydrophobic core scaling, was detected at low DNA concentrations, thus indicating an increased stability of the double-stranded conformation in Mg-DNA solutions as compared to the case of Na-DNA solutions. Some changes are detected in the behavior of the fundamental length scale pertaining to the single molecule DNA properties, reflecting modified electrostatic screening effects of the Odijk-Skolnick-Fixman type. All results consistently demonstrate that Mg$^{2+}$ ions interact with DNA in a similar way as Na$^{1+}$ ions do, their effect being mostly describable through an enhanced screening.
\end{abstract}

\maketitle
 
\section{Introduction}
\label{sec 1}
 
The biological functions of the highly negatively charged DNA are intimately coupled to the positive counterions that neutralize them (for a review see Refs. \ \onlinecite {Kornyshev2007,Cherstvy2011}). A particularly significant effect ubiquitous in biological environment is the {\em DNA condensation} in multivalent salt DNA solutions \cite{Rau246 92,Rau260 92,Pelta BJ96,Pelta JBC96,Plum88,Raspaud99}. In this case the presence of a multivalent ion atmosphere creates effective {\em attractive interactions} between nominally equally charged DNA molecules that can lead to toroidal aggregate formation, which seems to be the preferred morphology for high-density packaging of DNA \cite{Hud2005}. In fact, most vertebrate sperm cells contain DNA toroidal aggregates, each measuring about 100 nm in outside diameter, condensed by arginine-rich and thus highly charged proteins. Condensed DNA  aggregates seem to be relevant also for gene packing in bacteriophages and for their potential impact in artificial gene delivery \cite{Rudi2008}. While DNA condensation is also at least in part due to electrostatic interactions, it can not be explained within the mean-field Poisson-Boltzmann imagery \cite{Rudi2010}. A radical reformulation of the theory of electrostatic interactions is needed, based on the concept of "strong coupling" between the multivalent salt counterions and the charges on the DNA backbone, in order to understand the counterintuitive change in sign of electrostatic interactions between nominally equally charged bodies \cite{Rudibook}. While the general framework of the strong-coupling effects is well understood, there are different ways of its exact implementation that accentuate different facets of the interaction between the DNA backbone and the mobile multivalent ions in the bathing solution \cite{Cherstvy2011,Teif}. In order to elucidate and eventually differentiate between different theoretical approaches more detailed experiments on the effects of multivalent ions are needed. While the investigation of equilibrium properties of multivalent counterion DNA solutions are vigorously pursued \cite{Rau246 92,Rau260 92,Pelta BJ96,Pelta JBC96,Plum88,Raspaud99}, the concomitant elucidation of their dynamical properties leaves a lot to be desired. 

Dynamics of multivalent counterions would be in particular interesting to probe in order to reveal otherwise inaccessible features of the strong-coupling attractive interactions between DNA molecules \cite{Rudi2010}. However, not all multivalent ions in DNA solutions act the same and they are clearly differentiated in the way they affect the ds-DNA. Early light-scattering studies indicated that DNA collapse into compact structures occurs when $\approx90\%$ of the DNA phosphate charges are neutralized by condensed counterions \cite{Wilson79}. It also became clear that in fact very few cations induce ds-DNA condensation. Univalent and divalent cations, excluding transition-metal ions such as Mn$^{2+}$, Ni$^{2+}$, and Cu$^{2+}$  \cite{Hud2001}, do not condense ds-DNA even when present at very high concentrations, while almost all divalent cations condense the single-stranded DNA but not the ds-DNA.  On the other hand, alkaline-earth divalent cations, such as Mg$^{2+}$, Ba$^{2+}$ and Ca$^{2+}$, do condense triple stranded DNA with a more highly charged helix than the ds-DNA form, but not the ds-DNA \cite{Qiu2010}. Furthermore, counterions Mn$^{2+}$, Cd$^{2+}$, Co(NH$_3$)$^{3+}$, polyamines such as spermidine$^{3+}$, spermine$^{4+}$, polylysine$^{+}$, etc. do condense ds-DNA at finite concentrations. While electrostatics without doubt plays an important role in the DNA condensation mechanism \cite{Cherstvy2011}, it cannot be the sole factor affecting it as, e.g., Co(NH$_3$)$^{3+}$ is more efficient in condensing DNA than spermidine$^{3+}$, both being trivalent counterions. The best condensing agents appear to be those that bind into one of the DNA grooves \cite{best}. Based on these varied properties of multivalent counterions, it is thus clear that their dynamical behavior could provide an additional clue to the puzzles posed by DNA collapse.
 
In what follows, we thus plan to investigate the dynamics of multivalent counterions by using the dielectric spectroscopy (DS) to study DNA solutions at various DNA densities and salt content. DS \cite{Buchner2009} has been successfully used to probe the dynamics of aqueous polyelectrolyte solutions \cite{Nandi, Bordi2004}. The higher-frequency dielectric response located in high MHz-GHz region (near 17 GHz) is usually associated with the macromolecular hydration and the bulk dielectric response of water \cite{Hastedbook}.  On the other hand, the lower-frequency modes (between 100 Hz and 100 MHz) are dominated by the counterion atmosphere around each polyelectrolyte and this fluctuating charge cloud responds to the applied electric field \cite {Bordi2004}. We recently resolved the counterion response to external fields into two fundamental dissipative peaks that appear as fingerprints of two relaxation modes arising from the diffusive motion of counterions \cite{DNAPRL06, PRE2007, DNAEPL, Vuletic2010}: the dynamics of free diffuse counterions detected in the MHz frequency range probes collective properties of the polyelectrolyte solution, whereas the dynamics of condensed counterions in the kHz range probes the single-chain polyelectrolyte properties. It is noteworthy that on the latter time scale the polarization of counterions happen along an essentially stationary DNA chain. We have shown that the details of the counterion dielectric response depend on various parameters characterizing the polyelectrolyte chain and the polyelectrolyte solution as a whole, such as the length of the polyelectrolyte chain, its charge density and flexibility, and the concentration of polyelectrolyte chains, as well as the ionic strength of the added salt. 
 
DNA has been studied by DS since the early 1960s \cite {Sakamoto76, Mandel77, Bordi2004}. In the case of univalent cations such as Na$^{1+}$, the MHz relaxation in the semidilute regime of DNA solutions corresponds to the collective (de Gennes--Pfeuty--Dobrynin) correlation length as the defining property. This is true for DNA as well as for other polyelectrolytes, e.g., hyaluronic acid \cite{Vuletic2010}. Since the DS technique can be applied at very low polyelectrolyte concentrations, even below 10 g/L, it can be seen as complementing scattering techniques, such as SAXS and SANS, that are usually applicable only for much higher polyelectrolyte solution concentrations \cite{TomicMacromolecules}. Recent experiments by Salomon {\it et al.} \cite{VuleticMacromolecules2013} showed that the characteristic length scale probed in SAXS measurements corresponds to the characteristic length scale of the HF dielectric relaxation obtained by DS, thus confirming that both techniques probe the same correlation length of the semidilute DNA solutions. The specific features of DNA, not observed for other polyelectrolytes, are pronounced flexibility of short double-stranded DNA (ds-DNA) fragments \cite{DNAEPL,Yuan2008, Lee2010} and locally fluctuating regions of exposed hydrophobic cores of long DNA \cite{PRE2007}. The latter can be associated with the nucleation of DNA denaturation bubbles stemming from broken segments of several consecutive base pairs \cite{bubbles}. The kHz relaxation, on the other hand, reveals the single molecule properties such as the flexibility of a single DNA chain as described by its persistence length. Its variation with the solution electrostatic screening (Debye) length can be rationalized by the Odijk-Skolnick-Fixman (OSF) theory, applicable to rigid and semi flexible chains \cite{Odijk, Fixman}. 
 
In order to shed additional light on the behavior of ds-DNA in solutions of polyvalent salts we thus embark on a systematic study of the dielectric response of DNA solutions in multivalent salt bathing solutions. As a point of departure we choose the Mg$^{2+}$ cation. Our choice is based on the fact, {\sl vide supra}, that Mg$^{2+}$ does not induce any DNA condensation but nevertheless still affects DNA solution dynamics through enhanced screening, as will become evident as we proceed. This differentiation between the condensation effects and the screening effects is important in order to keep a clear track of the specificities of the multivalent ions. In subsequent studies we plan to move from the predominant screening effects to more complicated condensation effects as present with Mn$^{2+}$ or Co(NH$_3$)$^{3+}$ ions, elucidating their interrelation and prevalence.  Consistent with our previously elaborated methodology, we will probe the two dielectric response modes involving counterions in DNA solutions and attempt to rationalize our findings within the conceptual framework elaborated for the case of univalent salt DNA solutions and adapted specifically to the case of multivalent counterions.

\section{Materials and methods}
\label{sec2}        
 
Salmon testes lyophilized Na-DNA threads were purchased from Sigma-Aldrich. Previous gel electrophoresis measurements showed that the majority of DNA fragments were in the 2--20 kbp range, so that the estimated average DNA fragments were 4 $\mu$m long \cite{PhysicaB2012}. As for the Na-DNA solutions we were always in the semidilute regime \cite{PRE2007}. A crude estimate based on the de Gennes arguments gives the chain overlap concentration $ c^\ast$ of the order of 0.001 g/L, which is one order of magnitude below the lowest concentration found in our experiments and even if we take into account the shortest chains in our DNA solutions we estimate that $ c^\ast \approx 0.007$ g/L, which is still lower than the lowest concentration of DNA solutions in our measurements \cite{deGennes76, note1}. 
 
First we prepared the Mg$^{2+}$ salt of DNA in pure water and in solutions with added MgCl$_2$  salt. The reason for this is that we want to study DNA aqueous solutions containing only magnesium cations in order to avoid possible complications due to the presence of sodium cations \cite{Lyons64}. Thus Mg-DNA pure water solutions within concentration range $0.01$ g/L $\leq c \leq 4$ g/L were prepared by exhaustive dialysis according to the protocol as described in Ref. \ \onlinecite{noteprotokol1}. Then DNA solutions with different added salt ionic strengths were made. MgCl$_2$ was added to Mg-DNA water solution with a concentration chosen in such a way as to have the added salt ionic strength in the range $\text{0.003 mM} \leq I_{\mathrm{s}}\leq \text{4 mM}$ \cite{note2}. In addition, Mg-DNA solutions with concentrations in the range $\text{0.01 g/L}\leq c \leq \text{4 g/L}$ and with the added salt ionic strength $I_{\mathrm{s}}=$~0.3 mM and $I_{\mathrm{s}}=$~3 mM were prepared \cite{noteprotokol2}. Finally, and for the sake of comparison, Na-DNA solutions in the presence of a MgCl$_2$  buffer were also made according to protocols as above.

\begin{figure}
\resizebox{0.43\textwidth}{!}{\includegraphics*{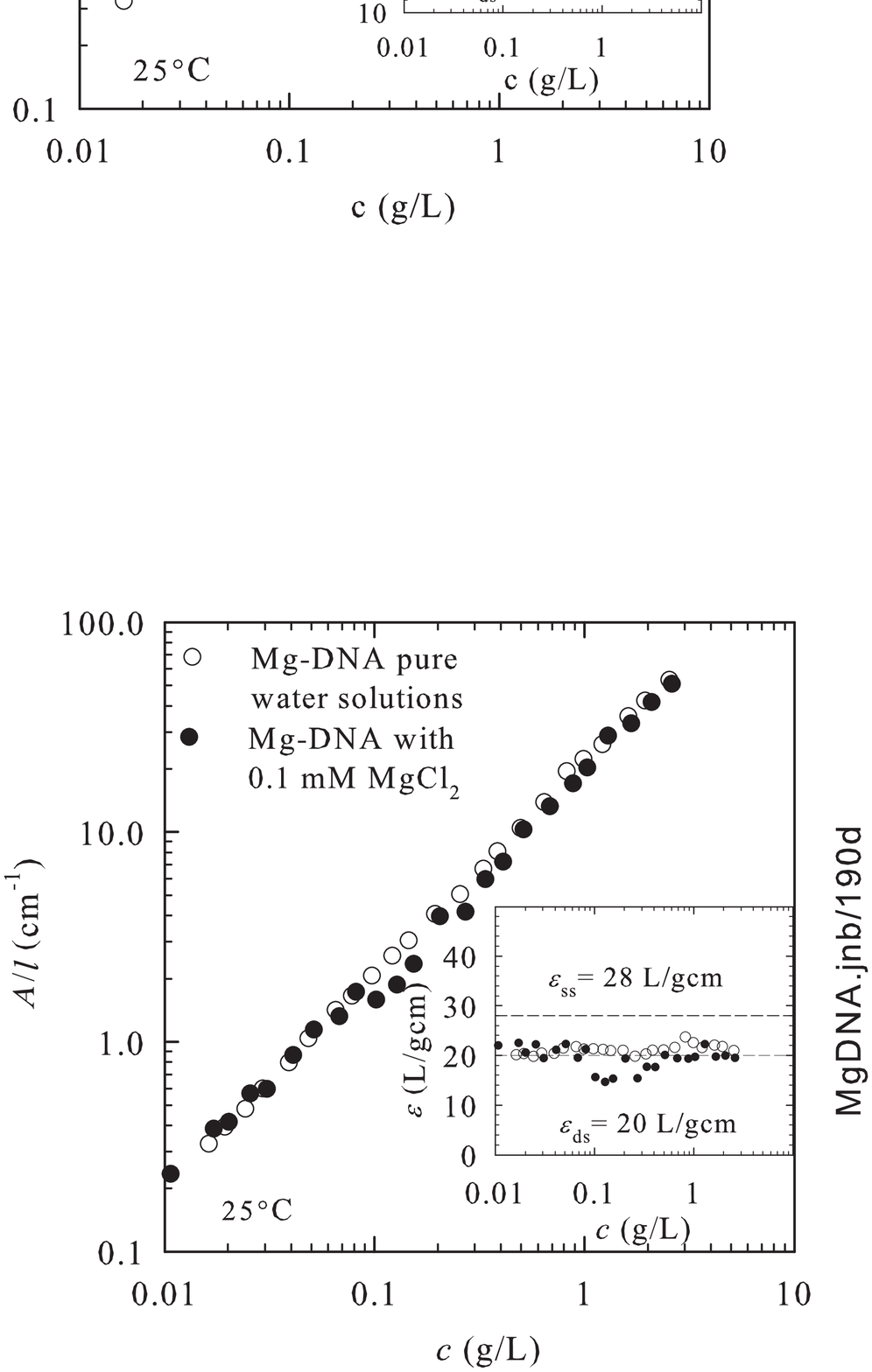}}
\caption{Normalized absorption ($A/l$, where $l$ is the light path length) of Mg-DNA pure water solutions (open circles) and of Mg-DNA solutions with added salt MgCl$_2$, $I_{\mathrm{s}} = 0.3$ mM (full circles) as a function of DNA concentration. Inset: extinction coefficient versus Mg-DNA concentration indicating double-stranded conformation in the whole concentration range.}
\label{fig1}
\end{figure}

UV spectrophotometry measurements of the DNA absorbance intensity at 260 nm were done in order to verify the nominal DNA concentration. The concentration was then determined assuming double-stranded conformation, implying the extinction coefficient at 260 nm equal to 20 L/gcm. The measured concentrations for DNA solutions in 1 mM of added salt were consistently smaller by about 20\% than the nominal ones, as found before \cite{ PRE2007}. This difference is due to water content not taken into account by the spectrophotometry approach. Throughout this paper we refer to these measured concentrations as the {\em DNA concentrations}. We have also used UV spectrophotometry in order to verify the stability of the double-stranded conformation of Mg-DNA solutions. The obtained data clearly showed that the ds conformation was fully preserved not only in added salt, but also in the pure water solutions in the whole range of studied concentrations (Fig.\ \ref{fig1}).

\begin{figure}
\resizebox{0.48\textwidth}{!}{\includegraphics*{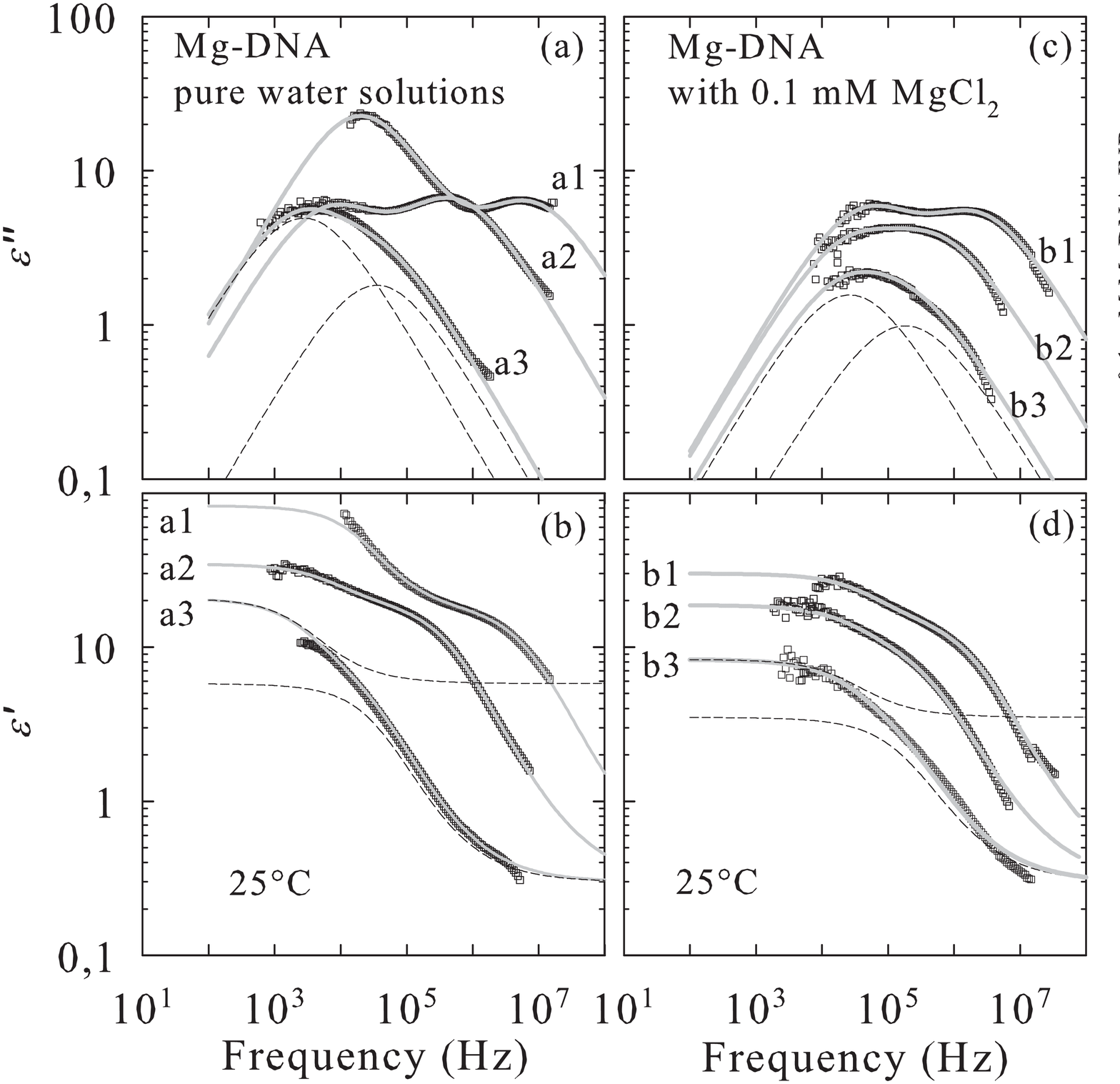}}
\caption{Imaginary ($\varepsilon^{\prime\prime}$) and real ($\varepsilon^\prime$) part of the dielectric function at $T=25^\circ$C of (a),(b) pure water Mg-DNA solutions and (c),(d) Mg-DNA solutions with added salt MgCl$_2$, $I_{\mathrm{s}}= 0.3$ mM for representative a1--a3 (4, 0.5, 0.022 g/L) and b1--b3 (2, 0.4, 0.08 g/L) DNA concentrations. The full lines are fits to the sum of the two Cole-Cole forms; the dashed lines represent a single CC form.}
\label{fig2}
\end{figure}
 
Dielectric spectroscopy measurements were performed at room temperature (25$^\circ$C) using a setup which consisted of a home made capacitive chamber with parallel platinum electrodes and a temperature control unit, in conjunction with the Agilent 4294A precision impedance analyzer operating in $\nu= 40$~Hz--110 MHz frequency range. The capacitive chamber enables reliable complex admittance measurements with reproducibility of 1.5\% of samples in solution with small volume of 100 $\mu$L and with conductivities in the range 1.5--2000 $\mu$S/cm. The chamber constant value is $l/S = 0.1042\pm 0.0008$~cm$^{-1}$, where $S = 0.98$~cm$^{2}$  is the effective electrode cross section corresponding to the sample of 100 $\mu$L and $l = 0.1021\pm 0.0001$~cm is the distance between the electrodes. Extrinsic effects, especially those due to free ions and electrode polarization, were removed by using the reference subtraction method. To this end, MgCl$_2$ solutions of different molarities were chosen for reference samples and measured in addition to DNA samples. This method gives reliable results up to the concentration of 2 g/L. At higher concentrations the influence of electrode polarization is sufficiently large both due to counterions and added salt ions, so that it cannot be subtracted in a satisfactory way. Because of that we were only able to determine the parameters of the high-frequency relaxation for concentrations below 2 g/L. Detailed measurement procedure and data analysis was given previously \cite{PRE2007, PhysicaB2012}. Since then an additional improvement in the data analysis has been implemented. In our previous method, the most time-consuming and difficult to automate task was finding the appropriate electrolyte solution to be used as a reference: simple algorithmic matching in most cases requires an additional manual adjustment. With this in mind we have written a custom fitting program in Python using the open source Enthought Tool Suite (Traits, TraitsUI, Chaco). It provides interactive visualization and matching of the background to sample spectra. Model fitting can then be applied immediately to the resulting dielectric spectra. In the case of two overlapping dielectric modes near the high-frequency limit of our experimental window, we additionally rely on fitting the model to the real part of conductivity. In particular, this improves the accuracy of the extracted mode width.

\section{Results}
\label{sec3}        

\begin{figure}
\resizebox{0.48\textwidth}{!}{\includegraphics*{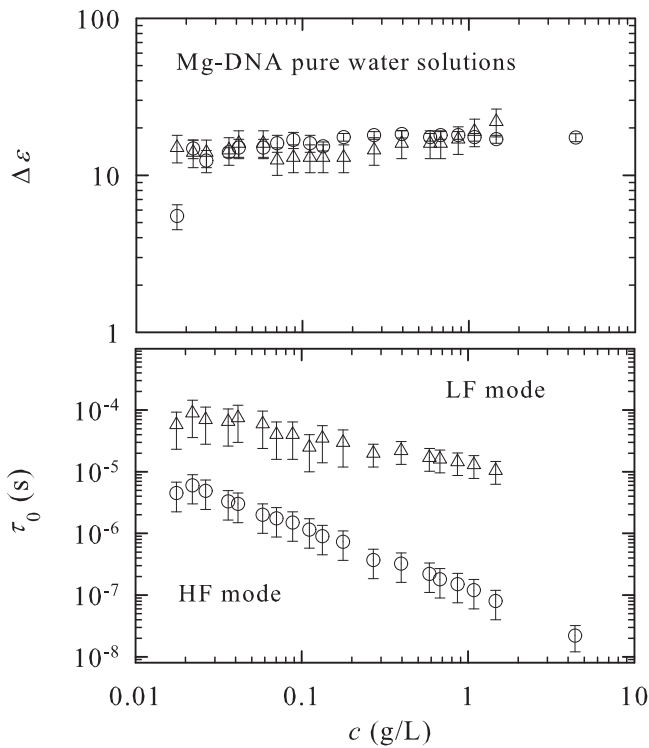}}
\caption{Dielectric strength ($\Delta\varepsilon$, upper panel) and mean relaxation time ($\tau_0$, lower panel) of Mg-DNA pure water solutions as a function of DNA concentration. Open circles and triangles stand for the HF and LF mode, respectively.}
\label{fig3}
\end{figure}
  
\begin{figure}
\resizebox{0.48\textwidth}{!}{\includegraphics*{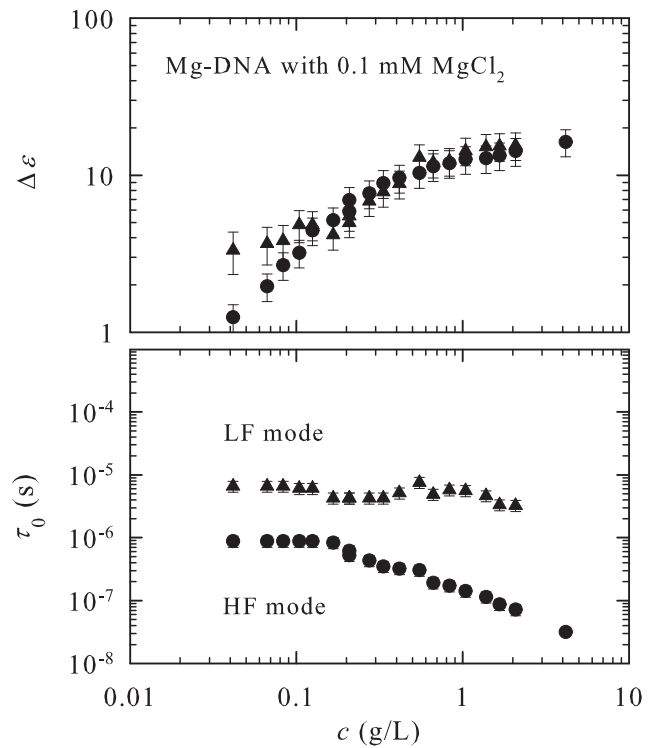}}
\caption{Dielectric strength ($\Delta\varepsilon$, upper panel) and mean relaxation time ($\tau_0$, lower panel) of Mg-DNA water solutions with added salt MgCl$_2$; $I_{\mathrm{s}} = 0.3$ mM as a function of DNA concentration. Full circles and triangles stand for the HF and LF mode, respectively.}
\label{fig4}
\end{figure}

Figure \ref{fig2} shows representative spectra for Mg-DNA pure water solutions and for solutions with 0.1 mM added salt. The complex dielectric spectra can be described by the sum of the two Cole-Cole functions 
\begin{equation}
\varepsilon(\omega)-\varepsilon_\infty
 = \frac{\Delta\varepsilon_{\mathrm{LF}}}{1 + \left(i \omega \tau_{0,\mathrm{LF}} \right)^{ 1-\alpha_{\mathrm{LF}} }}
 + \frac{\Delta\varepsilon_{\mathrm{HF}}}{1 + \left(i \omega \tau_{0,\mathrm{HF}} \right)^{ 1-\alpha_{\mathrm{HF}}}},
\label{Cole}
\end{equation}

where $\varepsilon_\infty$ is the high-frequency dielectric constant, $\Delta\varepsilon$ is the dielectric strength, $\tau_0$ the mean relaxation time, and $1-\alpha$ the symmetric broadening of the relaxation time distribution function of the low-frequency (LF) and high-frequency (HF) dielectric mode. 
The broadening parameter of both modes is typically $1 -\alpha \approx 0.8$.  The concentration dependencies of the dielectric strengths and mean relaxation times for pure water Mg-DNA solutions and solutions with added salt are shown in Fig.\ \ref{fig3} and Fig.\ \ref{fig4}, respectively.
 
From the measured mean relaxation time $\tau_0$ we first extract the characteristic length scales $L$ along which counterions diffuse following the applied ac electric field. We use the expression $\tau_0 \propto L^2/D_{\mathrm{in}}$, where $D_{\mathrm{in}}$ is the diffusion constant of counterions which is well approximated by the diffusion constant of bulk ions \cite{PRE2007, PhysicaB2012, Angelini2006}. Since we work with Mg-DNA aqueous solutions, we use the diffusion constant of Mg$^{2+}$ ions $D_{\mathrm{in}}=0.706\times 10^{-9}$ m$^2$/s \cite{textbook}. 
 
\subsection{HF mode}
\label{A}
 
\begin{figure}
\resizebox{0.48\textwidth}{!}{\includegraphics*{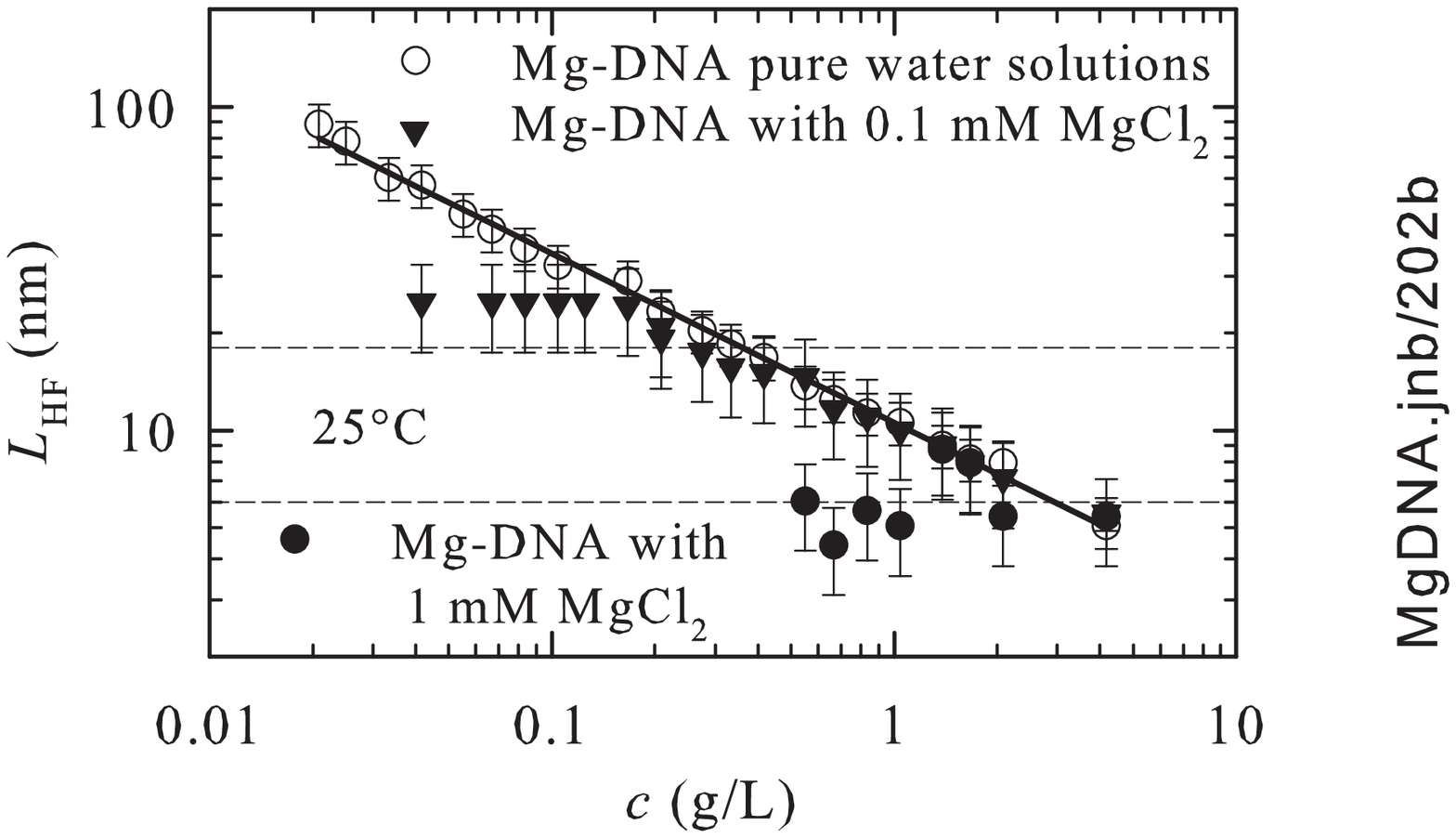}}
\caption{Characteristic length of the HF mode ($L_{\mathrm{HF}}$) for pure water Mg-DNA solutions (open circles) and for Mg-DNA solutions with added MgCl$_2$ salt; $I_{\mathrm{s}}= 0.3$ mM (full triangles) and 3 mM (full circles) as a function of DNA concentration ($c$). The full line is a fit to the power law $L_{\mathrm{HF}} \propto c^{-0.53\pm0.03}$. The dashed lines stand for the theoretically expected Debye screening lengths for the investigated $I_{\mathrm{s}}$ of added salt.}
\label{fig5}
\end{figure}

Our first important result concerns the characteristic length of the HF mode which probes the collective properties of DNA solutions. For pure water DNA solutions with Mg$^{2+}$  counterions, the characteristic length $L_{\textrm{HF}}$ follows the power law $L_{\textrm{HF}} \propto c^{-0.53\pm0.03}$ in an almost three-decades-wide range of concentration (Fig.\ \ref{fig5}). Thus, in the semidilute solution regime with divalent Mg$^{2+}$ counterions the de Gennes--Pfeuty--Dobrynin (dGPD) correlation length or the mesh size $\xi\propto c^{-0.5}$ \cite{deGennes76, Dobrynin} is revealed as the most relevant length scale. This is in distinction to the univalent Na$^{1+}$ counterions, where the mesh size $\xi$ characterizes the organization of DNA chains only at concentrations larger than 0.5 g/L, whereas at lower concentrations it is replaced by the exposed hydrophobic core scaling $c^{-0.33}$, corresponding to incipient denaturation bubbles \cite{ PRE2007,TomicMacromolecules}. As we  suggested previously, this may reflect the appearance of locally fluctuating denaturation bubbles due to the relatively weak effect of Na$^{1+}$ counterions that are unable to screen the repulsion between two DNA strands at these low concentrations \cite{DNAPRL06, PRE2007}. An increased stability of the double-stranded conformation detected in the case of Mg-DNA can thus be interpreted with diminished repulsion between two DNA strands. Both effects can be associated with the pure screening action of divalent magnesium counterions. These results thus suggest a much enhanced screening effect of magnesium counterions as compared with sodium, eventually yielding an increased stability of the ds conformation of DNA. A very similar conclusion has been reached by Lyons \al{}\ \cite{Lyons64} on the basis of their counterion activity coefficient data and assuming that binding is basically determined by the long-range electrostatic interactions, with site-specific binding playing only a marginal role. They have also found that the activity coefficient of Mg-DNA was five times smaller than the one for Na-DNA which indicates much stronger binding of magnesium cations to DNA.

With MgCl$_2$ as added salt, the behavior of $L_{\mathrm{HF}}$ remains unchanged (Fig.\ \ref{fig5}), so that again the dGPD correlation length is the only relevant length scale. This remains the case as long as the ionic strength of the Mg-DNA is larger than the ionic strength of the added salt. At lower DNA concentrations, the $L_{\mathrm{HF}}$ clearly levels off, with a limiting value close to the Debye screening length theoretically expected for this salt ionic strength. Qualitatively, this result is similar to the one found previously for Na-DNA water solutions with NaCl as the added salt. However, a significant change is found at the quantitative level indicating again an enhanced screening capability of DNA with Mg$^{2+}$ cations as compared to DNA with Na$^{1+}$ cations. Previously observed results indicate that NaCl added salt with the ionic strength only twice the ionic strength of Na-DNA prevails in the DNA solution screening and promotes the Debye screening length to the fundamental length scale \cite{DNAPRL06, PRE2007}. Conversely, the results displayed in Fig.\ \ref{fig5} show that adding of MgCl$_2$ to an ionic strength of six times the effective value of Mg-DNA is needed to overcome the intrinsic screening of Mg-DNA \cite{note4}.
 
\begin{figure}
\resizebox{0.48\textwidth}{!}{\includegraphics*{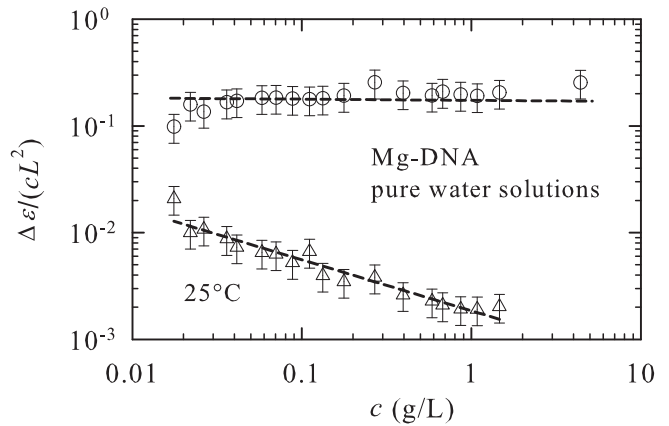}}
\caption{Normalized dielectric strength $\Delta\varepsilon /(c\cdot L^2)$ of the HF mode (open circles) and of the LF mode (open triangles) as a function of DNA  concentration $c$ for pure water Mg-DNA solutions. The dashed lines are guides for the eye.}
\label{fig6}
\end{figure}
 
Finally, our data enable an estimate of the number $f$ of oscillating counterions which is given by $\Delta\varepsilon/(c\cdot L^2)$ \cite{PRE2007}. Increasing DNA concentration leaves the fraction of free counterions participating in the HF relaxation process unchanged (Fig.\ \ref{fig6}); that is, it features qualitatively similar, concentration-independent $f$ as found for Na-DNA \cite{PRE2007}. This result validates the standard theoretical models which use the Manning-based definition of $f$ as the concentration-independent parameter.    

\subsection{LF mode}
\label{B}

\begin{figure}
\resizebox{0.48\textwidth}{!}{\includegraphics*{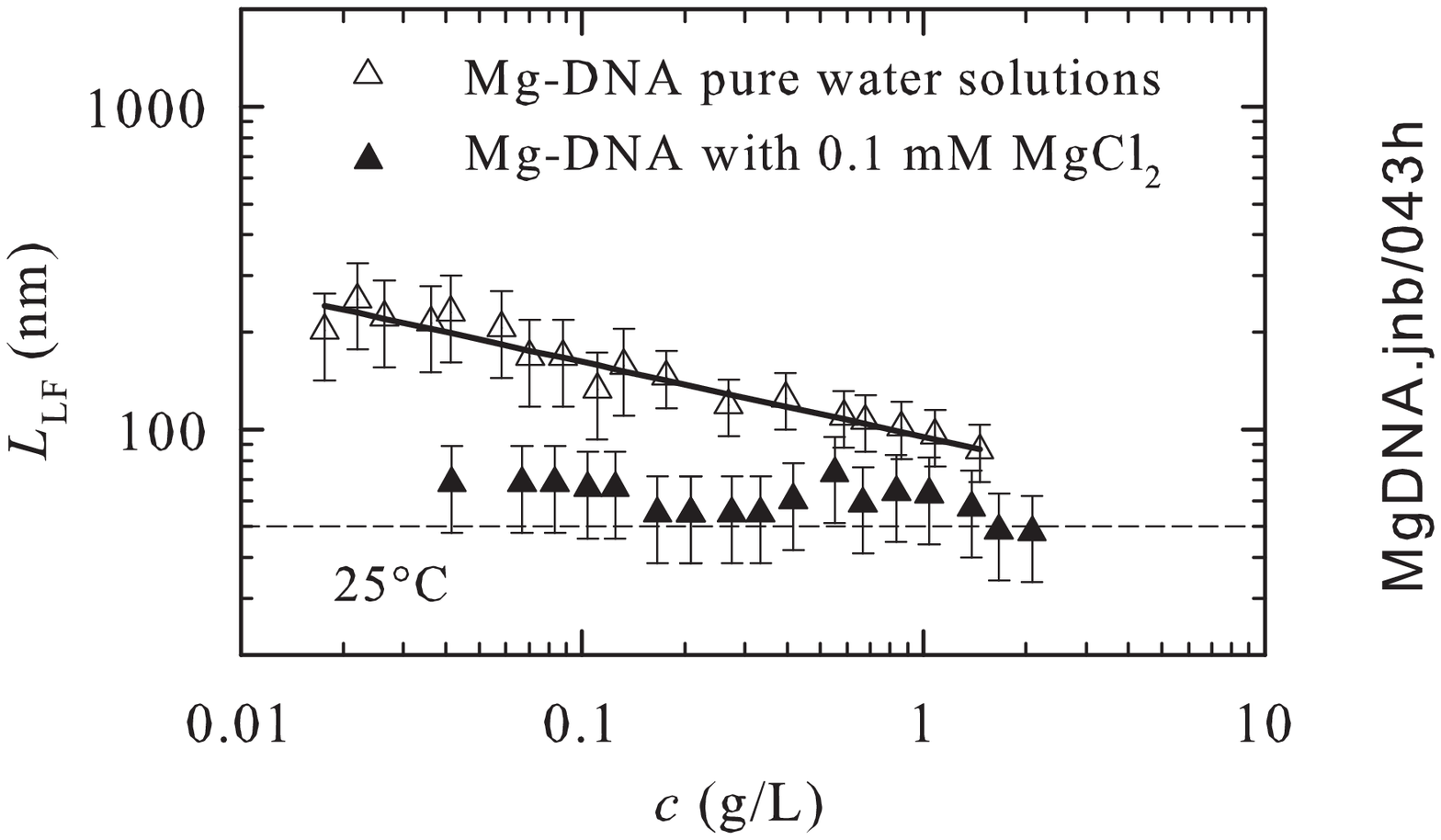}}
\caption{Characteristic length of the LF mode ($L_{\textrm{LF}}$)  for pure water Mg-DNA solutions (open triangles) and for DNA solutions with added MgCl$_2$ salt; $I_{\mathrm{s}}= 0.3$ mM (full triangles) as a function of DNA concentration ($c$). The full line is a fit to the power law $L_{\textrm{LF}} \propto c^{-0.23\pm0.02}$. The dashed line designates theoretically expected value for DNA structural persistence length $L_0 = 50$ nm.}
\label{fig7}
\end{figure}

The second important result concerns the LF mode which characterizes the single-chain properties. As for Na-DNA, we find the characteristic length scale $L_{\textrm{LF}} \propto c^{-0.23\pm0.02}$ to behave as predicted for a Gaussian chain composed of correlations blobs, that is $L_{\textrm{LF}} \propto c^{-0.25}$ (Fig.\ \ref{fig7}). However, in Mg-DNA this characteristic length scale is shorter by about 1.5 times than in the Na-DNA \cite{DNAPRL06, PRE2007}, an effect which can again be ascribed to the enhanced screening in Mg-DNA. This result is also in accord with the viscosity data which show that Mg-DNA is hydrodynamically shorter than Na-DNA \cite{Lyons64}. The enhanced screening in Mg-DNA is also reflected in an effective number of condensed counterions participating in the LF relaxation (Fig.\ \ref{fig6}) which, in contrast to the Na-DNA \cite{PRE2007}, decreases with an increase in DNA concentration. The latter result indicates that the concentration-independent Manning-based definition for the number of oscillating counterions is not valid in this case. A rather surprising result in comparison to Na-DNA data is then obtained in the presence of added MgCl$_2$ salt with  $I_{\mathrm{s}}= 0.3$ mM (Fig.\ \ref{fig7}). In this case, $L_{\textrm{LF}}$ stays approximately constant in the whole measured concentration range at the level of  $L_0 = 50$ nm, which suggests that in this regime $L_{\textrm{LF}}$ is proportional to the structural persistence length of DNA. 

\begin{figure}
\resizebox{0.48\textwidth}{!}{\includegraphics*{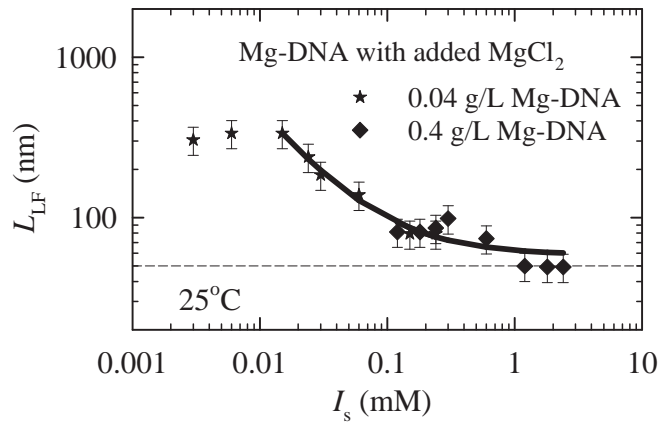}}
\caption{
Characteristic length of the LF mode ($L_{\textrm{LF}}$) for DNA solutions with varying added salt ($I_{\mathrm{s}}$) for two representative DNA concentrations: $c=0.04$~g/L (stars); $c=0.4$~g/L (diamonds). The full line is a fit to the expression $L_p = L_0 + a \cdot I_{\mathrm{s}}^{-1}$ with $L_0 = 57\pm5$~nm and $a =  4.4\pm1.2$~nm~mM.
}
\label{fig8}
\end{figure}

The dependence of $L_{\textrm{LF}}$ on the added salt ionic strength $I_{\mathrm{s}}$  is shown in Fig.\ \ref{fig8} for two selected Mg-DNA concentrations. The observed data follow the OSF behavior $L_p = L_0 + a \cdot I_{\mathrm{s}}^{-1}$, where $L_0$ is DNA structural persistence length and $a$  is the effective charge density of DNA \cite{Odijk, Fixman}. From the fit to the OSF expression we get $L_0 = 57\pm5$ nm and $a = 4.4\pm1.2 $nm$\,$mM \cite{note3}. While the value of $L_0$ close to $50$ nm is in accordance with standard expectations as well as experimental results for DNA structual persistence length \cite{Bloomfield00, Baumann97}, the value of the coefficient $a$ which describes the effective linear charge density is two times smaller than the value found in Na-DNA. A qualitatively similar result has been previously found by a magnetic birifrigence experiment \cite{MaretWeil}. 
It is noteworthy that both values of the coefficient $a$ are much smaller than the value expected by the standard OSF theory $a=\frac{\alpha^2}{z^2}32.4$ nm$\,$mM, where $\alpha$ is the degree of ionisation, $z$ is counterion valency \cite{note3}. There could be several reasons for this type of discrepancy, quite probably based on the limitations inherent in the OSF theory itself that completely disregards any ion-specific effects \cite{Ben-Yaakov2011}. Ion specificity is a compound effect and has many potential sources including polarizability, London-dispersion effects, and other nonpolar interactions that could play a significant role also in ion-DNA interactions \cite{Misra95}. While these effects can modify the ion-DNA interaction they are as a rule long range and cannot be viewed as short-range tight binding of ions to specific sites along the DNA. 

Figure \ref{fig8} also attests to the respective roles of intrinsic DNA counterions and ions from the added salt in ionic screening. As in the case of Na-DNA, the OSF model applies as long as the ionic strength of added salt is larger than the ionic strength pertaining to DNA. In the opposite limit the OSF behavior is replaced by the DNA self-screening. Again, comparing with the Na-DNA case,  an important quantitative difference is detected indicating that ten times stronger ionic strength of added salt is needed for Mg-DNA as compared to Na-DNA in order for the OSF behavior to prevail. The data for $c = 0.04$ and 0.4 g/L deviate from the OSF behavior for $I_{\mathrm{s}} < 0.01$~mM and for $I_{\mathrm{s}}< 0.1$ mM, respectively. At these low added salt values $I_{\mathrm{s}} < 0.04 I_{\text{Mg-DNA}}$, where $I_{\text{Mg-DNA}}=4c$ \cite{note4}, the intrinsic counterions become dominant in determining the behavior of $L_{\textrm{LF}}$ and indeed it attains the same value as in pure water Mg-DNA solutions (Fig.\ \ref{fig8}). On the other hand, Na-DNA experiments \cite{PRE2007} show that deviation from the OSF behavior takes place when $I_{\mathrm{s}} < 0.4 I_\text{Na-DNA}$, where $I_\text{Na-DNA}=3c$. In the case of divalent Mg counterions it thus appears that the screening from DNA and its counterions largely overpowers the screening effects of added salt, thus promoting the average size of the chain, as opposed to the OSF persistence length, to the fundamental length scale.

\section{Discussion}
 \label{sec4}        

It is important to note that our experiments were performed in the absence of any other kind of cation except Mg$^{2+}$. This presents an important point and allows for a straightforward interpretation when comparing our data with other published qualitatively similar results on DNA with magnesium cations (see \cite{Lyons64} and references therein). Specifically, a shorter statistical end-to-end distance was found by Foerster resonance energy transfer (FRET) experiments for the single-stranded DNA \cite{Chen2012}. In addition, the FRET measurements also showed that Mg$^{2+}$ is even 20--40 times more efficient in screening than Na$^{1+}$ in the case of single-stranded DNA. In a previous DS investigation, Na-DNA solutions with concentration 0.25 g/L were studied in the presence of different amounts of MgCl$_2$ so that the effects of increasing ratio Mg$^{2+}$/base pair were tracked \cite{Bonincontro2002}. Owing to the restricted frequency range in that work, only the dielectric relaxation in MHz range (that is HF) was detected. The data showed an increase of the correlation length $\xi$ and a decrease of the number of oscillating counterions $\Delta\varepsilon_{\textrm{HF}}/(c \cdot \xi^2)$ with increasing ratio Mg$^{2+}$/base pair. The authors of that study suggested that their results might be understood as a consequence of a strong site-binding of Mg$^{2+}$ so that they could not contribute to the dielectric relaxation \cite{Clement73}.  However, these DS data showing larger values of the correlation length for larger amounts of Mg$^{2+}$ ions can be easily understood within a framework where both sodium and magnesium cations can respond equally to applied ac electric fields. Namely, at the low Na-DNA concentration used in that work, the correlation length is short if compared with pure Mg-DNA due to the prevailing hydrophobic scaling with exponent 0.33 instead of the dGPD mesh size scaling with exponent 0.5. Evidently the increasing Mg$^{2+}$ content leads to a larger screening thus yielding larger values of the correlation length. Indeed, we have verified this scenario in our experiments on Na-DNA solutions with MgCl$_2$ added salt. Site-specific binding to DNA was also excluded by Raman data which indicated relatively weak interactions of Mg$^{2+}$ as well as of Na$^{1+}$  with both base and phosphate sites \cite{Duguid93}. 
Thus we are lead to conclude that the magnesium counterions interact with DNA prevalently via long-range electrostatic forces and that they respond to applied ac fields in the same way as the sodium cations.

\section{Conclusion}
\label{sec5}        
 
In conclusion, we have analyzed the behavior of magnesium counterions in ds-DNA semidilute solutions by means of dielectric spectroscopy measurements. Our results demonstrate that divalent magnesium cations significantly contribute to the DNA self-screening of  electrostatic interactions. Their screening efficiency is in fact anomalously large when compared to univalent sodium cation standard. However, with magnesium counterions there is as yet no net attraction between DNA segments that would be capable of inducing the full DNA collapse. Other types of multivalent counterions will be studied in order to probe these effects. 
 
As follows from our DS experiments, the effect of magnesium cations on the properties of DNA solutions can be characterized by a shortening of the statistical end-to-end distance of the DNA chains and a smaller effective linear charge density than for the Na-DNA, a clear indication of an ion-specific effect. The magnesium cations also assure the stability of the double-stranded conformation even in the limit of low DNA and added salt concentration by making the dGPD solution correlation length the most relevant fundamental length scale that describes the structural organization of DNA chains in solution. We also find that specific-site short-range strong counterion binding, sometimes assumed to be relevant for DNA counterion interaction, appears to not be particularly relevant for the interaction of magnesium cations with DNA in the studied concentration range of DNA and added salt. On the contrary, we confirm that the counterion-DNA interaction is primarily due to long-range electrostatic forces acting in a similar fashion as in the case of sodium counterions.  Further characterization of a denatured state of DNA in the presence of magnesium cations and other multivalent counterions is currently on the way.

\section*{Acknowledgments}
ICP-AES was performed by I.\ Ladan at the Croatian National Institute of Public Health. Discussions with T.\ Vuleti\'{c} and D.\ Rau  in the early stage of this work, as well as with  M.\ Tomi\'{c} are acknowledged. This work was supported by the Croatian Ministry of Science, Education and Sports under Grant No. 035-0000000-2836. R.P. acknowledges the financial sup\-port from the Slovene Agency for Research and Development (Grants No. P1-0055 and No. J1-4297).


\begin{thebibliography}
\bibitem{note1}
$c^\ast$ is given by the concentration where there is only one polymer molecule
in the volume of a polymer globule $c^\ast = \textrm{molecule mass}/V_{\mathrm{c}}$, where molecule mass is $N \cdot m_{\mathrm{bp}}$, and
$V_{\mathrm{c}} \approx L_{\mathrm{c}}^3 = N^3 \cdot a^3$; $m_{\mathrm{bp}}$ is a mass of a base pair $\approx 10^{-18}$ mg, $L_{\mathrm{c}}=N \cdot a$; $a=0.34\ \text{nm}$; $a$ is the monomer size.
\bibitem{noteprotokol1}
Stock solution was prepared by dissolving dry DNA in 25 mM of MgCl$_2$  for 48 h at $4^{\circ}$C. Next, dialysis starting from the stock DNA solution was performed in three steps: against 10 mM MgCl$_2$, against 1 mM MgCl$_2$, and finally against pure water. Each dialysis with two exchanges with ratio 1:500 was carried out during 48 h at 4$^\circ$C. Such a procedure was sufficient to remove all sodium ions as determined by ion chromatography.
\bibitem{note2}
Note that the ionic strength and molar concentration for MgCl$_2$ are related as $I_{\mathrm{s}}= 3c$.
\bibitem{noteprotokol2}
The procedure starting from the stock solution was applied with the final dialysis against 0.1 mM and 1 mM of MgCl$_2$, respectively.
\bibitem{note4}
Ionic strength of Na-DNA and Mg-DNA can be written as: $I_{\mathrm{DNA}} = (1/2) [z_p^2c_p+z_i^2c_i]$ where $ z_p c_p= 2c$, $ z_p c_p = z_i c_i$ and $ z_i c_i = 2c$. $z$ is valency, $c$ is the concentration of base pairs, while $c_i$ is the counterion concentration which for Na-DNA reads $c_i = 2c$, while for Mg-DNA: $c_i = c$. Thus it follows that the ionic strength of Na-DNA and Mg-DNA reads $I_{\mathrm{DNA}} = 3c$ and $I =  4c$, respectively.
\end{thebibliography}
\end{document}